# Comparative Analysis of Hand-Crafted and Machine-Driven Histopathological Features for Prostate Cancer Classification and Segmentation

Feda Bolus Al Baqain and Omar Sultan Al-Kadi*

King Abdullah II School of Information Technology, Artificial Intelligence Department, University of Jordan, Amman, Jordan

Email: fda8160332@ju.edu.jo (F.B.A.); o.alkadi@ju.edu.jo (O.S.A.)

*Corresponding author

*Abstract*—Histopathological image analysis is a reliable method for prostate cancer identification. In this paper, we present a comparative analysis of two approaches for segmenting glandular structures in prostate images to automate Gleason grading. The first approach utilizes a hand-crafted learning technique, combining Gray Level Co-Occurrence Matrix (GLCM) and Local Binary Pattern (LBP) texture descriptors to highlight spatial dependencies and minimize information loss at the pixel level. For machine-driven feature extraction, we employ a U-Net convolutional neural network to perform semantic segmentation of prostate gland stroma tissue. Support vector machine-based learning of hand-crafted features achieves impressive classification accuracies of 99.0% and 95.1% for GLCM and LBP, respectively, while the U-Net-based machine-driven features attain 94% accuracy. Furthermore, a comparative analysis demonstrates superior segmentation quality for histopathological grades 1, 2, 3, and 4 using the U-Net approach, as assessed by Jaccard and Dice metrics. This work underscores the utility of machine-driven features in clinical applications that rely on automated pixel-level segmentation in prostate tissue images.

## I. Introduction

Prostate Cancer (PCa) is considered one of the foremost predominant and critical types of cancer among men, and it is considered the fourth main reason of cancer death in men [1]. An estimation of 1,806,590 new cancer cases and 606,520 cancer deaths are projected to occur in the United States [2]. PCa symptoms are considered moderate developing and non-lethal, while if left untreated, some develop and spread rapidly with fatal outcomes. This relates to the associated unclear symptoms. The difficulty in diagnosing PCa lies in the requirement of multiple procedures. Among these, is detecting the presence of cancer regions in tissue by examining prostate tissue biopsy by pathologists. A Gleason grading score, which determines the severity of cancer grade, is conventionally assigned to the examined tissue [3].

Early detection and precise diagnosis of Prostate Cancer (PCa) plays a vital role in the treatment of this disease. Physicians tend to discuss with patients symptoms they may have to find out if prostate problem persists. In order to decide whether a patient may have PCa, diagnosis tests are required. The traditional process consists of digital rectal examination and prostate-specific antigen blood test, followed by transrectal ultrasound guided biopsy [4]. The biopsy is recommended by clinicians if abnormality signs were found in the initial tests, as biopsies are considered the gold-standard for diagnosis. Extracted tissues will be stained with biomarkers, usually Hematoxylin-Eosin (H&E), a widely used method for tissue slides preparation (Fig. 1). Then stained tissue is analyzed under microscope for spotting suspicious areas, while increasing progressively the level of detail. Finally, samples are classified into a score from 1 to 5 according to Gleason grading [5]. The Gleason grading system is considered the standard system for PCa diagnosis. It is a standard endorsed by the World Health Organization and adopted worldwide by the pathologists [6].

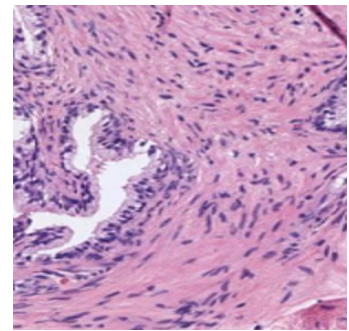

Fig. 1. Tissue stained with Hematoxylin and Eosin (H&E) staining.

Prostate tissue consists of gland units, the main components of the prostate gland unit are stroma, lumen, epithelial nuclei, epithelial cytoplasm, and blue mucin.

Fig. 2 illustrates the prostate gland component. Examining the glandular architecture for the prostate tissues, and the use of Gleason grading system helps the analysis and the description of cancer cell abnormalities as it involves assigning specific scores—ranging from 1 to 5—to each tissues based on the severity of cancer. These grades verify the aggressiveness of PCa, where benign tissues are described using grades 1 and 2, and grade 5 advanced.

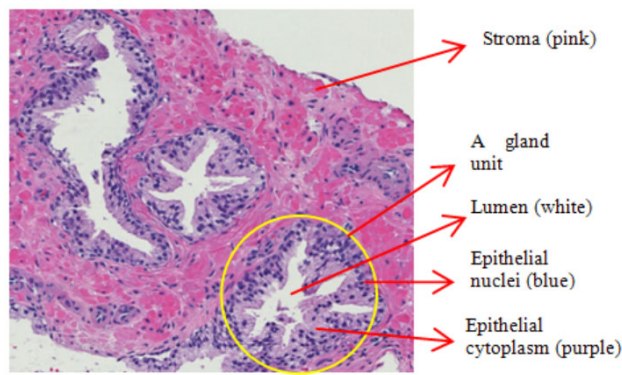

Fig. 2. Prostate gland components.

In benign tissues (i.e. grades 1 and 2), glands appear as large single separated units, densely packed and having large branchy lumen components and thick gland boundaries with prominent nuclei. In this respect, Gleason grade 3 is the most frequent case of carcinoma. It is characterized by the irregular glandular patterns and the invasion of small glands into the stroma, also has small, circular lumen and thin nuclei boundaries [7], Furthermore, the pathological tissue exhibits a reduced cell density in the layer of epithelial nuclei. However, in Gleason grade 4, the glands start to lose their architecture; glands tend to fuse nearby gland we can notice the loss of normal gland units and structure become ill-defined. It is obvious that this grade does not have a well-separated gland unit with separate lumen, and well-defined epithelial cell layers on the boundary. Nuclei distribute uniformly instead of forming well-defined boundaries as in benign patterns and glands are poorly defined. Multiple glands are mixed to form a mass of glands containing multiple lumen components. Finally, Gleason grade 5 is characterized by the total loss of gland structure. It is rarely found in men whose prostate carcinoma is diagnosed early [8]. This grade is commonly simply differentiated by the presence of a large variety of scattered nuclei on the stroma, since its features are not significantly distinguishing. For that reason, grade 5 is not considered in this research.

Since image texture serves a key role in identifying objects or specific regions of interest in an image, this work investigates how texture features of prostate images can help in improving PCa diagnosis and accurate Gleason grading. For this purpose, statistical methods for examining texture that consider spatial relationship of pixels using Gray Level Co-occurrence Matrix (GLCM), and Local Binary Pattern (LBP) was adopted and tested [9]. In addition, the Convolutional Neural Network (CNN) will be used for finding patterns to recognize an object in an image and the preferences of eliminating the need for manual feature extraction. The process of allocating Gleason scoring to a histopathology image is subjective and time consuming due to tissue complexity and physician interpretation, which often leads to elevated levels of intra and inter-observer variability. To this end, the need for increasing reproducibility of the grading process and to save pathologist time arises. Although there are different techniques that have been suggested so far, the development process of a reliable algorithm for PCa diagnosis is still an open problem. Thereby, there is a need for investigating the effectiveness of extracted image features for a reliable and robust PCa diagnosis system, especially for segmenting Gleason grade 2 and 3.

Accordingly, our work aims to develop a model for accurate diagnosis of PCa using histopathology images and Gleason scoring. Gland features, which exhibit significant variation in size, shape, and color intensity of different tissue components, have potential for classification of tissue images. We adopt texture and color features of the prostate images and quantify them using GLCM and LBP features. Segmenting prostate tissue images into gland and stroma objects based on extracted features from tissue glands is a challenging task that can be summarized in three stages: feature extraction, classification, and segmentation. These stages rely on image processing, machine learning, and deep learning techniques. Discriminating between grades 3 and 4 is especially difficult because gland morphology tends to be altered, leading to loss of structure and eventually reduced effectiveness of automated diagnosis. Our work investigates the effectiveness of hand-crafted features versus machine-learned features for classifying and segmenting gland stroma regions. We employ a traditional hand-crafted learning approach using GLCM and LBP with sliding windows to extract spatial relationships of pixel features at different displacements and angles, as well as an automated feature extraction process using deep-learning techniques in a specific encoder-decoder architecture based on a U-Net model. In the next stage, we employ various machine learning algorithms to classify gland from stroma components using hand-crafted extracted features from the previous stage. Finally, we perform a segmentation process to achieve a good separation of glands.

We introduce a simple online decision support tool designed to assist pathologists in distinguishing between normal and cancerous tissues. The web tool, accessible at http://193.188.66.253/webapps/home/index.html, can accurately segment glands and stroma objects in prostate tissue, providing faster and more reliable second opinion support. By offering an objective measure, the automated tool has the potential to overcome the inherent subjectivity and dependence on pathologists' level of expertise. The basic focus of this work is to improve the reliability and validity of diagnosis, especially in rural clinics with limited consultants and infrastructure, and thus represents a valuable contribution to the field of diagnostic pathology.

This research work presents several contributions, including:

1. A comparative analysis of two different feature extraction methods, namely hand-crafted and machine-driven, for the classification of gland and stroma image patches.
2. Evaluation of both hand-crafted features (based on GLCM and LPB) and machine-driven features (based on U-Net autoencoder architecture) for pixel-level semantic segmentation.
3. Comparison of the accuracy and training time of different feature extraction approaches with recent works.
4. Implementation of a web tool using the best pixel-level segmentation approach.

The following is the organization of the paper: In Section II, the related literature is presented in detail. Moving forward, Section III discusses the technical approach and the data used in the study. In Section IV, the experimental results are presented, and the main findings are interpreted. Finally, Section V provides a conclusion to the work.

## II. RELATED STUDIES

Many studies over the past years have examined and investigated different approaches for medical image segmentation using both machine and deep learning approaches. Methods performing prostate image segmentation using hand-crafted feature extraction techniques and pixel-level base features extraction using U-Net architecture will be discussed.

Rezaeilouyeh *et al.* [10] propose the use of a Discrete Shearlet Transform (SHT) for feature representation. A histogram of Shearlet coefficients was implemented as feature for classification, SHT was compared with other filtering techniques like wavelets and Gabor filter and achieved a promising accuracy result of 89% in cancer diagnosis between benign and malignant prostate tissue according to Gleason grading. However, [6] targeted to have an automatic classification from biopsy images for PCa. A multi-classifier system was proposed, based on tissue descriptors to explain textural features of the histopathology images. Six different methods were applied for the purpose of feature extraction; the first three were measured from the Quaternion Wavelet Transform coefficients, quaternion ratios, and the histograms of multiresolution local binary patterns and were used with the Bayesian classifier. However, the other features were based on discrete Haar wavelet transform, color fractal dimension, and morphometric characteristics of the tissue and were used with Support Vector Machine (SVM) classifiers. The simulations of the system were carried out on a dataset of 71 images of H&E-stained prostate tissue and achieved 98.89% correct classification rate. Another work by Ren *et al.* [11] for prostate gland segmentation was proposed and applied on different staining 18 images, to distinguish the Gleason 3 and 4 score in images from various structural graphic, region-based nuclei segmentation was used to in order to eliminate the need for lumen as prior information and to get individual gland. Progression from grade 3 to grade 4 was measured by calculating a gland shape variation score. The work assumes that the number of local maximal points within the glandular region represents the number of glands. Results achieved for precision, recall, and F1 were 94%, 60% and 70% respectively.

Other works were based on local structural modeling to implement automatic Gleason grading, using segmented tissue of the component, for each sub-image, a local structure feature was extracted [12]. A lumen-nuclei co-location feature was used to model sub-graphs features, learned as bags-of-words features for each labeled grade sample. Codebook and 3-class SVM classifier were used to obtain the structural similarity between sub-graphs in unlabeled images and the representative sub-graphs. The 300 H&E stained prostate histopathology images were used for testing and the average grading accuracies of 91%, 76%, and 65% were obtained on Grade 3, 4 and 5 samples respectively. In a similar manner, a computer aided diagnosing system for automatic grading of PCa tissue, was proposed by Ali *et al.* [13], which was based on discrete wavelet packet decomposition, where it was used to divide the image into all sub-band by providing a predefined number of levels. From each sub-band of the prostate tissue images four statistical features based on GLCM were extracted and Multiclass SVM achieved 92% accuracy.

In order to compare various feature extraction algorithms, Öztürk and Akdemir [14] used a multi-feature extraction algorithm and applied them on histopathologic images and compared the results using different classifiers. Feature matrices were extracted from cut image parts using Gray-Level Co-Occurrence Matrix (GLCM), Local Binary Pattern (LBP), Local-Binary Grey Level Co-Occurrence Matrix (LBGLCM), Gray Level Run Length Matrix (GLRLM), and Segmentation-based Fractal Texture Analysis (SFTA) as feature extraction techniques. Then they were classified using SVM, KNN, Linear Discernment Analysis (LDA) and Boosted Tree. The obtained feature matrix by SFTA algorithm produces the more successful result compared to other algorithms, and among classifiers Support Vector Machine (SVM) and Boosted Tree algorithms were the more effective ones.

In recent years, the use of convolutional neural network models has been explored to learn features directly from the image to avoid manual segmentation. In 2017, Zhou *et al.* [15] proposed the employment of Deep Neural Networks. Their work focused on strong classification to discriminate between Gleason grade 3+4 and 4+3, the method combined the extracted features from data engineering as well as those features that were learned automatically by the deep neural network. A method for optimizing color decomposition was developed for Hematoxylin density extraction. *K*-means clustering was used to extract the tumor part, in a way that they only focus on tumor part for that the subsequent learning and classification part. Finally, convolutional neural network classifiers trained on Gleason 7, the accuracy based on 368 slides of whole-slide images was 75%.

The architecture of the U-Net model was proposed by Ronneberger *et al.* [16] they extend the architecture of fully convolutional networks by adding a relative symmetric up-sampling path to down-sampling path, creating a U-shaped network architecture that can be worked with fewer training images. The architecture provides satisfactory results, it achieved an average IOU (Intersection Over Union) 77.5% on cell segmentation. U-Net was developed where Li *et al.* [17] developed multi-scale U-Net to predict four tissue classes at once (i.e. stroma, benign, grade 3, and grade 4). Mean Jaccard and 10-fold cross validation were used as evaluation metric to compare their proposed multi-scale U-Net with the stander U-Net and pixel wise CNN. The outperforming result was achieved compared with other models, providing a mean Jaccard of 65.8% across 4 classes, namely stroma, Gleason 3, Gleason 4, and benign glands, and 75.5% for 3 classes, namely stroma, benign glands, and PCa. Lately, U-Net has proven to be the most prominent deep network for medical image segmentation, therefore, Ibtehaz and Rahman [18] and Kalapahar *et al.* [19] proposed new modification on the standard U-Net architecture by adding residual block connection, a type of configuration in convolutional filters with skip-additive connections to improve the efficiency in segmenting biomedical images. Ibtehaz and Rahman [18] develop a novel architecture MultiResU-Net by adding as we have mentioned the residual block connection and introduce the 1×1 convolutional layer. They also greatly reduce memory requirement by factorizing the bigger, more demanding 5×5 and 7×7 convolutional layers, using a sequence of smaller and lightweight 3×3 convolutional blocks. They compare their result with the stander U-Net on five public datasets for medical image, using Jaccard index as evaluation metric and it shows improvement in performance of 10.15%, 5.07%, 2.63%, 1.41%, and 0.62% in using MultiResU-Net over U-Net. As well as Kalapahar *et al.* [19] proposed the Residual U-Net, and compare their performance with other deep networks (e.g. Fully Convolutional Networks, the SegNet and the stander U-Net architecture) and applying the deep learning segmentation models on full gradation of cancerous patterns in prostate biopsies, this dataset related to 96 patients, with a total of 182 prostate biopsies belong to them. Their proposed model achieved a pixel-level Cohen's quadratic kappa of 0.52 in the test data. Additionally, the work of Vacacela and Benalcázar [20] proposed prostate segmentation for the central gland and peripheral zone. The work evaluated and compared between two models applied on 2D semantic segmentation, the first model used encoder-decoder architecture on the global and local U-Net architecture, while the second model used encoder-classifier based on VGG16 pre-trained network, the experiments showed better performance for the former model when applied to Ductal Carcinoma as compared to the latter model that used a per-trained network. Comelli *et al.* [21] compared three U-Net model types: U-Net, efficient neural network, and efficient residual factorized convNet on prostate patient who went through MRI (Magnetic Resonance Imaging) examination, then applied resizing and data augmentation on the resulting MRI images, finally *k*-fold cross validation was applied and the result shows a Dice similarity coefficient of 90.89%.

In recent work by Liu *et al.* [22], a study was conducted on multi-parametric Magnetic Resonance Imaging (mpMRI) as a noninvasive alternative for Prostate Cancer (PCa) detection and characterization. They developed a Mutually Communicated Deep Learning Segmentation and Classification Network (MC-DSCN) based on mpMRI for prostate segmentation and PCa diagnosis. The primary goal was to design an MC-DSCN that jointly performs segmentation based on pixel-level information and classification based on image-level information. The proposed architecture transfers mutual information between segmentation and classification components, facilitating each other in a bootstrapping manner. The results showed Intersection over Union (IOU) percentages increasing from 84.5% to 87.8% and 83.8% to 87.1% for two different datasets, respectively. Furthermore, Gavade *et al.* [23] explored four deep learning architectures for mpMRI-based PCa segmentation and classification. The architectures included Semantic DeepSegNet with ResNet50, DeepSegNet with Recurrent Neural Network (RNN), U-Net with RNN, and U-Net with Long Short-Term Memory (LSTM). Their findings indicated that the combination of U-Net and LSTM achieved the best segmentation and classification results.

From the previous literature review, it is shown that histopathological image analysis is vital for accurate detection and diagnosis of PCa. However, a significant research gap exists in comparing the performance of hand-crafted features and machine-driven features, particularly in differentiating between normal and cancerous tissues at early stages. This work aims to fill this gap by conducting a comprehensive comparative analysis of these feature extraction approaches for PCa segmentation and classification. By investigating the effectiveness of hand-crafted features that capture domain-specific knowledge and machine-driven features that reveal complex patterns, this research seeks to provide insights into the optimal approach for accurate classification and segmentation. Furthermore, the development of an online web tool as a reliable second opinion for pathologists in regions with limited access to specialized expertise is a key goal, which will benefit communities with under-resourced healthcare settings.

## III. MATERIALS AND METHODS

### A. Dataset

To investigate the effectiveness of hand-crafted features that capture domain-specific knowledge against machine-driven features that uncover complex patterns, we conducted experiments using the dataset provided by García *et al.* [24]. We selected this dataset due to its inclusion of whole-slide images encompassing both healthy tissue and tumor prostate areas, making it highly compatible with our case study's pre-processing steps. The dataset's whole-slide images are well-suited for the techniques we employ, ensuring effective application and evaluation of our methods. Furthermore, the dataset facilitates comparative analysis with other studies and includes annotations by expert pathologists, ensuring the accuracy and reliability of ground truth labels. This high-quality annotation is critical for training and validating our models, contributing to robust and reproducible results. The dataset comprises 35 whole-slide images, with 17 representing healthy tissue and 18 representing tumor tissue, collected from 25 patients. The diversity of samples can enhance the generalizability of our work, where expert pathologists from the Hospital Clínico Universitario de València annotated the tissue images.

In order to prepare the data for analysis, a preprocessing step was performed. This involved identifying the region of interest using a bounding box and eliminating irrelevant pixel information (as illustrated in Fig. 3(a)). To enhance resolution and local information, sub-images of reduced size were created using a sliding window protocol (as shown in Fig. 3(b)). The resulting bounding box was divided into patches measuring 1024×1024 pixels, equivalent to an optical magnification of 10×. Patches containing less than 5% of tissue pixels, which provide minimal useful information, were discarded (as depicted in Fig. 3(c)). The final dataset consisted of 3,195 benign glands (as shown in Fig. 3(d)), 3,000 cancerous glands of Gleason grade 3, and 3,200 artifacts (i.e., false glands).

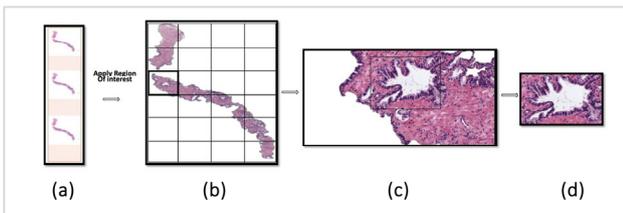

Fig. 3. Dataset preparation [25], (a) example of a whole-slide image; (b) region of interest from which we perform the sliding window protocol; (c) sub-image of 1024 × 1024 pixels from which we address the Gland Candidate; (d) Gland Candidate.

### B. Preprocessing Prostate Tissue

The pre-processing steps involved initially dividing the histopathology images into gland stroma patches. Subsequently, the input patches were resized to 384×384 pixels to ensure compatibility with CPU memory constraints. A total of 280 patches were collected, consisting of 125 stroma patches and 155 gland patches. For the hand-crafted feature extraction experiment, the gland stroma image patches were converted into grayscale color models.

### C. Feature Extraction

After the completion of the pre-processing steps on the histopathology prostate images, the image patches are prepared for the subsequent feature extraction stage. In this stage, we applied hand-crafted techniques to extract twelve features from the patches. To reveal the relation between pixel neighbors and minimize information loss, a sliding window method was employed, varying the displacement and angle, and utilizing the gray level co-occurrence matrix for one approach and local binary pattern feature extraction method for the other approach. It is important to note that these approaches differ in terms of the specific techniques employed for feature extraction where all the experiments were conducted using MATLAB 2020a (Mathworks, Inc).

#### 1) Gray level co-occurrence matrix

The spatial distribution of gray values plays a crucial role in defining texture characteristics, and deriving statistical features from these distributions has long been recognized as an early technique in the field of image processing. Haralick introduced the concept of co-occurrence matrices for extracting texture features, which remains one of the most widely used methods for capturing second-order statistical properties [26]. The process of extracting texture features involves two main steps: first, the computation of the co-occurrence matrix, followed by the interpretation of texture features based on spatial relationships. The Gray Level Co-occurrence Matrix (GLCM) is a two-dimensional histogram that represents the distribution of gray levels for pairs of pixels with a fixed spatial relationship. The GLCM is computed using a displacement vector defined by its distance ($\delta$) and orientation ($\theta$) (Gadkari, 2004). In our work, Fig. 4(a) illustrates an example of a GLCM with a distance ($\delta$) of 1 and a horizontal direction ($\theta = 0°$), representing the nearest horizontal neighbor relationship. Additionally, Fig. 4(b) shows the co-occurrence matrix offsets implemented in our work.

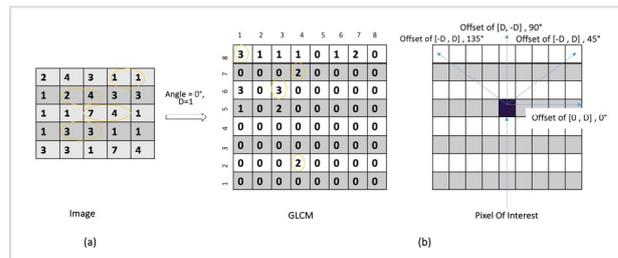

Fig. 4. Example of gray level co-occurrence matrix method on an image patch. (a) using offset of [0 1], (b) offsets implemented to create each extracted hand-crafted features.

#### 2) Local binary pattern

The Local Binary Pattern (LBP) technique is a first-order neighborhood method used for extracting local features from images by evaluating the intensity variations between the central pixel and its neighboring pixels. This approach aims to identify diverse patterns by assigning binary labels to each pixel based on its relationship with the center pixel. Specifically, for each neighbor, a 0 bit is assigned if its pixel value is smaller than the central pixel, whereas a 1-bit value is assigned if it is equal to or greater than the central pixel value. These binary bits are then concatenated in a clockwise order, and the binary representation of the central pixel is replaced with its corresponding decimal value. The resulting histogram of LBP labels, representing the frequency of occurrence of each label within an image region, serves as a texture descriptor for the image of interest [27], as illustrated in Fig. 5.

#### 3) Convolutional neural networks

In recent years, the remarkable advancements and exceptional performance of deep learning have captured the attention of researchers, establishing it as the preferred approach for numerous medical image analysis problems. These include tasks such as image denoising, semantic segmentation, and classification. Semantic segmentation, in particular, is a specialized image processing technique that involves the division of an image into distinct regions possessing similar characteristics, such as colors, textures, or other features. When objects sharing common properties are assigned the same label based on their semantic meaning, it is referred to as semantic segmentation. Undoubtedly, this technique holds significant importance and continues to be an active area of research within the field of computer vision.

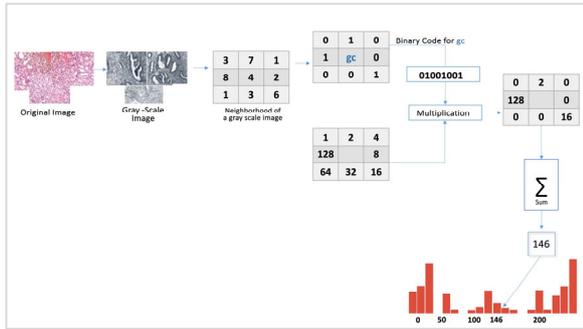

Fig. 5. Example of local binary pattern method using a 3×3 neighborhood threshold by central pixel value.

The task of semantic segmentation revolves around the assignment of each pixel in an image to a corresponding object class, effectively labeling every pixel with a specific category. To achieve this, the present work employed deep learning models based on the widely adopted U-Net architecture. U-Net is an encoder-decoder network architecture extensively utilized in medical image segmentation, as will be further discussed.

### D. Deep Convolutional Neural Network for Biomedical Image Segmentation

The histopathological images were segmented using the classical U-Net, a widely employed architecture for biomedical image segmentation. This architecture, depicted in Fig. 6, is constructed with meticulous design principles that contribute to its effectiveness.

The U-Net architecture comprises two main sections: the encoder and the decoder. The encoder serves as a contracting path and employs a series of progressive convolutions to extract distinctive features from the input image. This process involves two consecutive 3×3 convolutional layers, followed by a 2×2 max pooling operation with a stride of 2. This operation is repeated four times, enabling the extraction of increasingly complex features. It is noteworthy that the number of filters in the convolutional layers is doubled as the down-sampling operations progress deeper into the architecture, as depicted in Fig. 6. Connecting the encoder to the decoder are two additional 3×3 convolutional operations [18]

On the other hand, the decoder constitutes an extending path that incorporates a sequence of up-sampling operations followed by 2×2 convolutions. The decoder initiates the process by up-sampling the feature map through a 2×2 convolution operation, effectively reducing the number of feature channels by half. Subsequently, a series of two 3×3 convolutional operations is performed, repeated four times, successively decreasing the number of filters by half at each stage. Finally, a 1×1 convolution operation generates the segmentation map. Throughout the U-Net architecture, all convolutional layers, except for the final one, employ the Rectified Linear Unit (ReLU) activation function, while the last convolutional layer utilizes the sigmoid activation function [28].

One notable characteristic of the U-Net architecture is the extensive use of feature channels in the up-sampling path. This enables the propagation of contextual information from the contracting path to lower-resolution layers. Consequently, more precise localization is achieved in the expanding path, resulting in a symmetrical and U-shaped architecture [29].

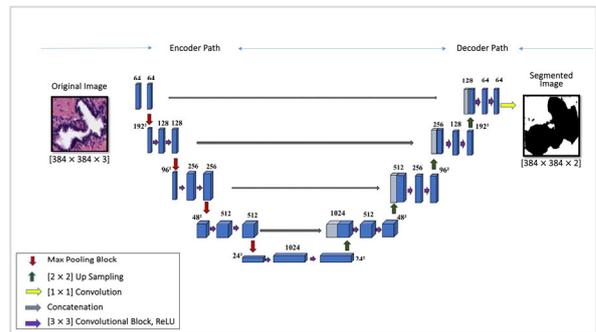

Fig. 6. Applied U-Net architecture for prostate image segmentation.

### E. Interactive Web Tool Design

The proposed model in this study encompasses multiple stages, each contributing to the overall process of prostate tissue segmentation. These stages include data collection, image preprocessing, hand-crafted feature extraction utilizing GLCM at various offsets, hand-crafted feature extraction using LBP at different radii, machine learning-based classification of the extracted features, fusion of features extracted from GLCM and LBP, feature selection of the hand-crafted features, image segmentation, and ultimately semantic segmentation employing the U-Net architecture. Fig. 7 illustrates the application of prostate tissue segmentation through this comprehensive process.

To enable histopathologists to obtain a second opinion on image segmentation, an interactive web tool has been developed. This user-friendly tool facilitates the uploading of histopathological prostate images for online image segmentation. The web tool's semantic segmentation functionality relies on the best segmentation model identified in the experimental results section of this work.

To tackle the classification and segmentation of gland and stroma objects, we explored five commonly used classifiers for hand-crafted feature extraction, as found in the existing literature. Specifically, Support Vector Machine (SVM), *k*-Nearest Neighbor, Bagging, Random Forest, and Naïve Bayes classifiers were utilized. For the semantic segmentation task, we employed the U-Net architecture as a classical convolutional network specifically designed for biomedical image segmentation. The U-Net model addresses the challenge of limited training data in medical image segmentation. The purpose is to achieve competitive segmentation results even with a smaller training dataset compared to traditional CNNs. Upon request authors should be prepared to send relevant documentation or data in order to verify the validity of the results. This could be in the form of raw data, samples, records, etc. Sensitive information in the form of confidential proprietary data is excluded.

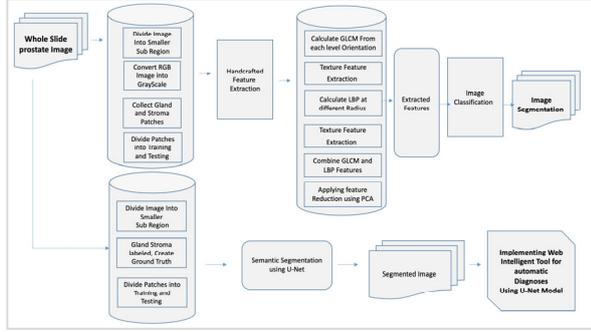

Fig. 7. Pipeline diagram of the interactive web tool design for automatic diagnosis of prostate cancer.

### F. Performance Assessment

Following the feature engineering extraction, selection, and implementation of various models for gland stroma image segmentation, the subsequent step involves evaluating the effectiveness of the model using different evaluation metrics on the test dataset. In this work, we employ accuracy as an evaluation metric for classification problems using a confusion matrix to compare the performance of different machine learning classifiers. Accuracy is defined as the ratio of correctly identified examples in all classes, as expressed by the equation:

$$Accuracy = \frac{TP + TN}{TP + TN + FP + FN} \qquad (1)$$

here, TP represents the number of true positives, TN represents the number of true negatives, FP represents the number of false positives, and FN represents the number of false negatives.

To assess the quality of the segmentation at the pixel level for each image, we utilize two evaluation indices: Jaccard index and Dice index. These indices measure the similarity between the set of pixels marked as ground truth ($X$) and the set of pixels segmented as glandular structures ($Y$). Both indices produce scores ranging from 0 to 1, where a score of 1 indicates a perfect segmentation [30]. Particularly, the Dice and Jaccard index are commonly used metrics to assess the similarity between ground truth and segmented images in semantic segmentation tasks. The Dice index quantifies the overlap between two sets and can be expressed as:

$$Dice_{index} = \frac{2 \times TP}{2 \times TP + FP + FN} \qquad (2)$$

where, TP represents the number of true positives, FP represents the number of false positives, and FN represents the number of false negatives. The Jaccard index, also known as the Intersection over Union (IoU) or Jaccard

similarity coefficient, measures the similarity and diversity of sample sets and can be expressed as:

$$Jaccard = \frac{TP}{TP + FP + FN}. \qquad (3)$$

For semantic segmentation architectures, additional metrics are commonly used to evaluate performance. These include the global accuracy, mean accuracy, mean Intersection over Union (IoU), weighted IoU, and mean boundary F1 (BF) score. The global accuracy metric calculates the ratio of correctly classified pixels to the total number of pixels in an image, irrespective of class. It is defined as:

$$Global_{Accuracy} = \frac{1}{\sum_i^n TP_i + FP_i} \qquad (4)$$

where, $TP_i$ represents the number of pixels predicted to belong to class $i$, and $FP_i$ represents the total number of pixels of class $i$ in the dataset. The global accuracy is expressed as a percentage. The mean accuracy metric computes the ratio of correctly classified pixels in each class to the total number of pixels, averaged over all classes:

$$Mean_{Accuracy} == \frac{1}{n}\sum_i^n \frac{TP_i}{TP_i + FP_i} \qquad (5)$$

where, $n$ is the number of different classes. The mean IoU metric calculates the average Intersection over Union for all classes in the dataset:

$$Mean_{IoU} = \frac{1}{n}\sum_i^n \frac{TP_i}{TP_i + FP_i + FN_i} \qquad (6)$$

where, $FN_i$ represents the number of pixels predicted to belong to class $i$ but are not assigned that class by the ground truth or predictor. The weighted IoU metric measures the average IoU of all classes, weighted by the number of pixels in each class:

$$Weighted_{IoU} = \frac{\sum_i^n w_i \times IoU_i}{\sum_i^n w_i} \qquad (7)$$

where, $w_i$ is the number of pixels in class $i$.

The mean BF score evaluates the similarity between the predicted and ground truth class boundaries, taking into account a pixel tolerance distance. The mean BF score is calculated as:

$$Mean_{BF\_Score} = \frac{1}{N}\sum_{k=1}^N \frac{2 \times Precision_k \times Recall_k}{Precision_k + Recall_k} \qquad (8)$$

here, $N$ represents the total number of images in the dataset. The precision ($Precision_k$) and recall ($Recall_k$) values between the predicted and ground truth class boundaries, given a pixel tolerance distance, are determined using the following equations:

$$Precision_k = \frac{TP_k}{TP_k + FP_k} \qquad (9)$$

$$Recall_k = \frac{TP_k}{TP_k + FN_k} \qquad (10)$$

where, $TP_k$ represents the number of true positives for class $k$, $FP_k$ represents the number of false positives for class $k$, and $FN_k$ represents the number of false negatives for class $k$. Among the evaluation metrics discussed, the mean Intersection over Union (IoU) is the most commonly used metric. It is regarded as a more precise metric compared to global accuracy since it penalizes false positive predictions.

## IV. EXPERIMENTAL RESULTS AND DISCUSSION

In this section, we present a detailed explanation of the conducted experiments along with their corresponding results. The experiments are organized into three main categories: firstly, hand-crafted feature extraction utilizing GLCM and LBP for image classification, followed by the segmentation process, and ultimately, the application of semantic segmentation using the U-Net architecture.

### A. Image Classification using GLCM

In this work, we employed the GLCM approach to calculate the co-occurrence matrix, allowing us to gather valuable information about the orientation of gland and stroma structures. For each offset and color image, we extracted a total of eight distinctive features. The utilization of different offset values enabled us to obtain diverse sets of hand-crafted features for each gland and stroma candidate. To evaluate the effectiveness of our approach, we conducted classification experiments employing cross-validation techniques, specifically $k$-fold and holdout methods. The entire process of image classification and segmentation using GLCM features is depicted in Fig. 8.

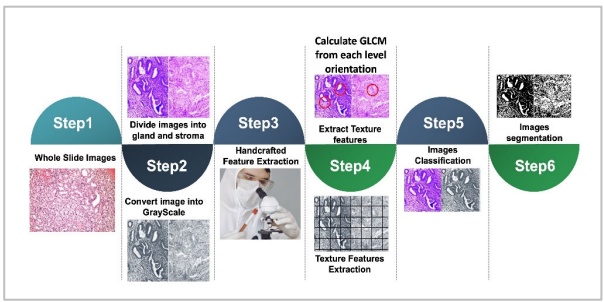

Fig. 8. Steps of image classification and segmentation using hand-crafted gray level co-occurrence matrix features.

To validate our results, we extracted hand-crafted features using the GLCM method from a dataset comprising 280 patches, including 125 stroma patches and 155 gland patches. We partitioned the histopathology images into gland and stroma patches, and subsequently extracted the GLCM features with different offsets. Specifically, we considered a $\delta$ of 1 for $\theta$ values of 0°, 45°, 90°, and 135°, and $\delta$ of 2, 4, 8, and 16 for all angles. The extracted features included Contrast, Entropy, Inverse Difference Moment, Correlation, Standard Deviation, Mean, and Angular Second Moment. These features were then utilized as inputs for five different classifiers: SVM, KNN, Bagging Trees, Naïve Bayes, and Random Forest.

The results indicate that the SVM classifier, with a $\delta$ of 1 and $\theta$ value of 0° using holdout validation (60% training, 40% validation), as well as the K-Nearest Neighbors (KNN) classifier, with a $\delta$ of 1 and $\theta$ value of 90° using 10-fold cross-validation, $\delta$ of 2 and $\theta$ value of 0° using 20-fold cross-validation, and $\delta$ of 4 and $\theta$ value of 0° using holdout validation (60% training, 40% validation), achieved the highest accuracy. We then employed the classifier that yielded the best accuracy on our testing patches to perform image classification and assign the testing patches to either the gland or stroma class. After applying the trained model to 104 testing patches, the SVM classifier achieved an accuracy of 99%, while the KNN classifiers achieved accuracies of 98% ($\delta = 1$, $\theta = 90°$), 96% ($\delta = 2$, $\theta = 0°$), and 98% ($\delta = 4$, $\theta = 0°$), respectively. Hence, based on the best accuracy results, SVM with a $\delta$ of 1 and $\theta$ value of 0° was determined as the optimal choice.

### B. Image Classification Using LBP

In this step, we employ the LBP approach to calculate texture primitives and gather local spatial information about the orientation of gland and stroma structures. For each radius and color image, we extract four distinctive features. By utilizing different radius values, we can obtain a unique set of hand-crafted features for each gland and stroma candidate. To evaluate the performance of our approach, we conduct classification experiments using cross-validation techniques, including $k$-fold and holdout methods.

The second approach for hand-crafted feature extraction involves the utilization of LBP on a total of 280 patches, comprising 125 stroma patches and 155 gland patches. Prior to model training, we perform the necessary preprocessing step of dividing the image into gland and stroma patches. LBP features are computed at radii of $r = 1$, $r = 2$, $r = 4$, $r = 8$, and $r = 16$, enabling the extraction of first-order features such as mean, standard deviation, skewness, and kurtosis. These features are then used as inputs for five different classifiers: SVM, KNN, Bagging Trees, Naïve Bayes, and Random Forest. All the experiments were conducted using MATLAB 2020a.

Our objective is to assess the performance of the trained model with higher accuracy on our testing patches, enabling image classification and assigning the patches to either the gland or stroma class. After applying the trained model to 104 testing patches, the results reveal that the SVM classifier achieved an accuracy of 95.1%. This accuracy serves as an indication of the model's capability to accurately classify the testing patches and distinguish between gland and stroma structures.

### C. Image Classification with Feature Reduction

For the creation of the Principal Component Analysis (PCA) space, we utilized a set of 280 training patches, consisting of 125 stroma patches and 155 gland patches. PCA was applied as a feature reduction technique on the twelve selected hand-crafted features from the GLCM and LBP methods that exhibited the best results. Specifically, GLCM features were obtained at ($\delta = 1$, $\theta = 0$), and for LBP, the radius value of 1 was employed.

The goal of PCA is to identify the directions of maximum variance in high-dimensional data and project it onto a new subspace with an equal or reduced number of dimensions compared to the original space. In our case, PCA was applied to the training features. Following the training process, three features with the highest variance were retained, with variances per feature listed in the following order: contrast (72.8%), correlation (18.5%), and energy (5.2%). The accuracy achieved by the model after PCA feature reduction was 96.4%. A summary of the results obtained from the various methods applied to the testing patches of histopathology prostate gland stroma is illustrated in Fig. 9.

From the experimental results, it is evident that the performance of the GLCM approach alone surpasses the performance of the combined GLCM+LBP approach when incorporating PCA. Furthermore, the SVM classifier consistently demonstrated superior performance compared to other classifiers when utilizing a 20-fold cross-validation strategy.

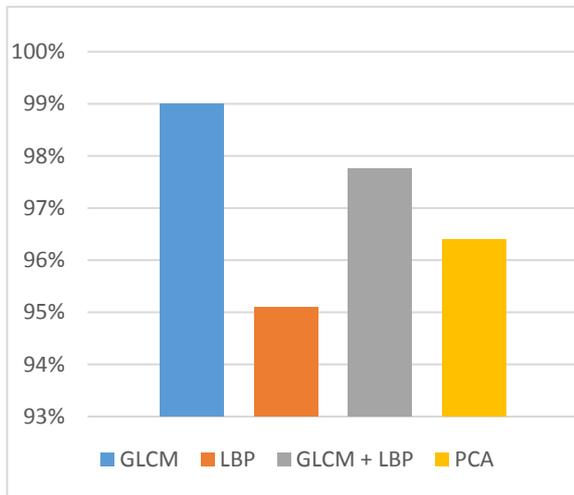

Fig. 9. Overall classification accuracy results for applied hand-crafted feature extraction methods.

### D. Image Segmentation using Individual and Combined Hand-Crafted Features

We compare the performance of prostate image segmentation using individual hand-crafted features, specifically focusing on gland and stroma segmentation. We investigate four model architectures: GLCM, LBP, GLCM+LBP, and PCA as a feature reduction technique. A comprehensive evaluation of segmentation results is conducted, comparing the performance of these different models.

To perform image segmentation, we adopt a sliding window protocol to analyze patches containing both gland and stroma components. GLCM and LBP features are extracted for each pixel in the image by applying a sliding window with a size of 35×35 pixels, suitable for images sized at 384×384 pixels. This approach yields a total of 147,456 features for an image of this size. The approximate inference time for each model is also measured, and it took approximately 2 minutes to complete the image segmentation process. The steps involved in image segmentation using combined hand-crafted features are depicted in Fig. 10.

To validate the effectiveness of the different hand-crafted models, we employ four distinct testing images to assess the performance of the models in gland and stroma image segmentation. Furthermore, we analyze the segmentation results using various evaluation metrics. Fig. 11 illustrates the original image, annotated image with pink annotations representing the stroma in the ground truth image, and gray annotations representing the gland. It also displays the segmentation results obtained using the GLCM, LBP, GLCM+LBP, and feature reduction models, respectively.

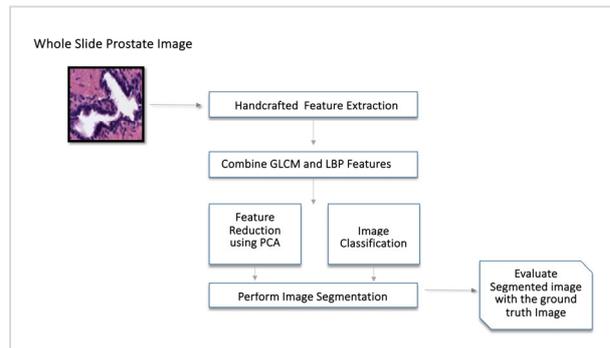

Fig. 10. Steps of image segmentation using hand-crafted features of both Gray Level Co-occurrence matrix and local binary pattern methods.

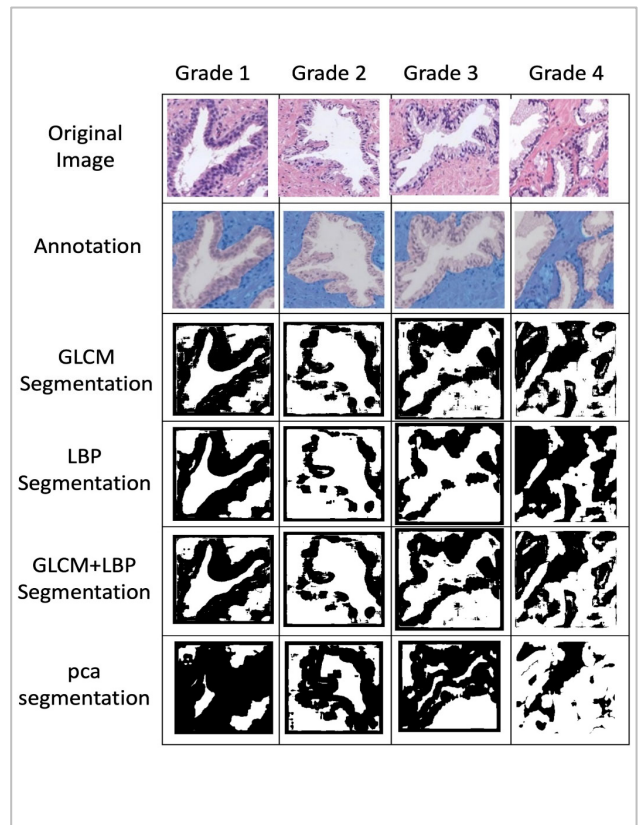

Fig. 11. Segmentation results using hand crafted models.

### E. Semantic Segmentation Using U-Net

Prostate image segmentation plays a key role in image-guided intervention. However, the lack of clear boundary,

and the huge variation of shape and texture between the images from different patients make the task particularly challenging. To overcome these problems. In this work, we utilize an upgrade of Fully Convolutional Network (FCN) for semantic segmentation, the U-Net architecture. Since deep learning simplified the process to perform semantic segmentation and achieve impressive results as previously shown, the U-Net was designed to obtain accurate segmentation with a few training images. The proposed model can effectively detect the gland stroma prostate region, where the task of semantic segmentation is to predict individual pixel values whether they belong to a specific interest region – in our case belong either to gland or stroma component.

### 1) Data set preparation

The input of the network includes two types of data: images to be learned from and the mask coordinates of the desired labeled objects of the image for the learning process (Gland and stroma patches and their prelabeled image). We labeled the image using the Image Labeler in MATLAB 2020a. A sample labeled image can be seen in Fig. 12 that was used for testing our U-Net training model. To perform a controlled training procedure, the architecture was trained with 60 epochs.

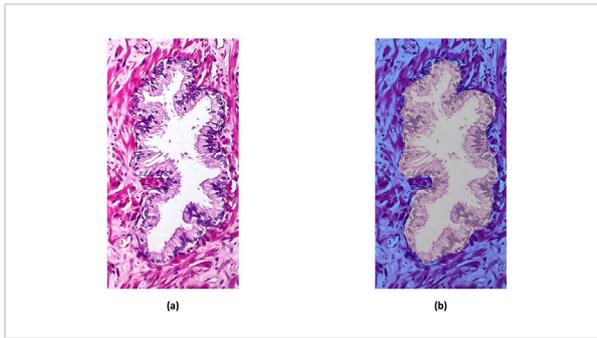

Fig. 12. Histopathological image showing gland and stroma: (a) Unlabeled image (b) Labeled image.

### 2) Training process

In this experiment, our dataset comprises images captured from various glands and stroma patches. We utilized a consistent dataset consisting of 200 Red, Green, Blue (RGB) images, each with fine pixel-level labeling. Among these images, 150 were designated for the training set, along with their corresponding target masks, while the remaining 50 images constituted the testing set.

For training the U-Net model, we employed a pixel-wise Softmax function for the final segmentation, which was combined with a cross-entropy loss function. This loss function analyzes each pixel individually, comparing the class predictions to our one-hot encoded target vector, which represents the labeled image. Since the class labels are nearly balanced, we selected cross entropy as our loss function. However, the Dice coefficient would be a suitable alternative in cases of class imbalance.

During the training process, the approximate training time was around fifteen hours, considering the specific environment settings employed. After training, the model achieved an accuracy of 94%.

### 3) Evaluating training network

The segmentation results achieved using the U-Net architecture for a representative image are presented in Fig. 13. Despite having a small training dataset, the U-Net model demonstrated superior performance in gland stroma image segmentation. The accuracy of the U-Net model was evaluated using testing images, and the approximate inference time for the model was measured, taking approximately 1 minute to perform patch image segmentation.

Various metrics were adopted to evaluate the segmentation results on the testing set, as shown in Table I, where the accuracy and Intersection over Union (IoU) segmentation values for each Gland and Stroma class are displayed. Additionally, the Mean Boundary F1 Score is provided.

The qualitative results demonstrate the capability of the proposed U-Net architecture to accurately segment prostate images into gland and stroma classes. To validate our trained model, we tested it on four different image patches, and the results are depicted in Fig. 13.

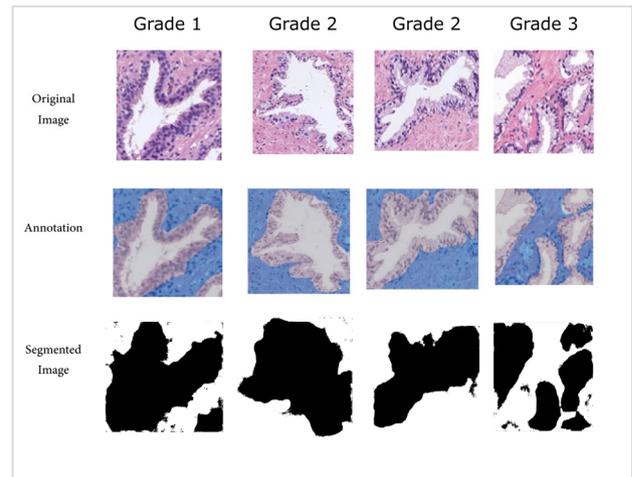

Fig. 13. Semantic segmentation result using the U-Net trained model.

TABLE I. ACCURACY AND IOU SEGMENTATION VALUE FOR EACH GLAND AND STROMA CLASS

| Semantic Segmentation | | Accuracy | IoU | Mean$_{BFScore}$ |
|---|---|---|---|---|
| U-Net | Gland | 0.91% | 0.80% | 0.40% |
| | Stroma | 0.95% | 0.91% | 0.81% |

### F. Performance Evaluation

To compare the image segmentations produced by the previous techniques with the labeled ground-truth images, two commonly adopted segmentation metrics, namely the Jaccard index and Dice index, were utilized. These metrics directly measure the generalization ability of the models in segmenting regions of interest on unseen hold-out data, specifically the gland and stroma segmentation of the prostate image patches. Both measures were calculated on the resampled image with a size of 384×384.

Table II provides a detailed overview of the segmentation results obtained in this study for grades 1, 2, and 3. The Jaccard and Dice indices demonstrate higher segmentation accuracy for the U-Net model. The results indicate that, in terms of segmentation performance, the automated feature extraction using the deep learning U-Net model, originally developed for biomedical image segmentation, outperforms hand-crafted features.

In contrast, a related work by [25] showed that hand-crafted features combined with an SVM classifier yield superior results compared to the deep learning VGG19 network in classifying benign vs. pathological (grade 3) images. The SVM classifier achieved accuracies of 88% compared to VGG19, which achieved 81% accuracy using 5-fold cross-validation to evaluate model performance. In their work, four different families of descriptors were employed for feature extraction, including morphological, fractal analysis, texture descriptors using GLCM and LBP, and contextual features. The SVM classifier applied to the GLCM and LBP features achieved accuracies of 99% and 95.1%, respectively, using 20-fold cross-validation. It should be noted that in their work, the last convolutional neural network in VGG19 was fine-tuned specifically for gland feature learning. However, the transfer learning approach with the generic VGG-19 network, trained on natural images, may not be as effective as utilizing the U-Net model, which was initially optimized for medical image segmentation and achieved an accuracy of 94% in our study. Fig. 14 visually compares the results obtained in our proposed work with those of [25].

Overall, the evaluation metrics demonstrate the superior performance of the U-Net model in our segmentation task, underscoring the importance of leveraging deep learning approaches specifically designed for medical image analysis.

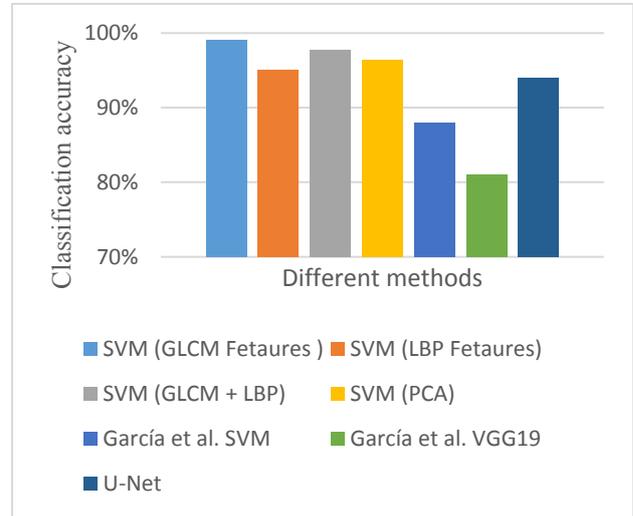

Fig. 14. Accuracy Comparison between our proposed work and [25].

TABLE II COMPARING IMAGE SEGMENTATION QUALITY FOR HAND-CRAFTED AND MACHINE-LEARNED MODELS

| Segmented Image | GLCM | | LBP | | GLCM+LBP | | PCA | | U-Net | |
|---|---|---|---|---|---|---|---|---|---|---|
| | Jaccard Index | Dice Index | Jaccard Index | Dice Index | Jaccard Index | Dice Index | Jaccard Index | Dice Index | Jaccard Index | Dice Index |
| Grade 1 | 0.30 | 0.46 | 0.35 | 0.52 | 0.33 | 0.49 | 0.19 | 0.19 | **0.85** | **0.92** |
| Grade 2a | 0.25 | 0.32 | 0.28 | 0.35 | 0.19 | 0.32 | 0.21 | 0.21 | **0.80** | **0.91** |
| Grade 2b | 0.24 | 0.35 | 0.34 | 0.48 | 0.35 | 0.52 | 0.39 | 0.39 | **0.82** | **0.90** |
| Grade 3 | 0.30 | 0.47 | 0.25 | 0.40 | 0.31 | 0.47 | 0.29 | 0.29 | **0.78** | **0.88** |

*G. Discussion*

The primary objective of this study was to explore the feasibility of developing both manual and automated computer models for histopathological prostate image segmentation. The goal was to investigate whether image patches could be utilized to mimic the pathologist's procedure and aid in accurately defining the gland and stroma components of histopathological prostate images, thereby assisting in the diagnosis of PCa. Various models were evaluated, including hybrid features using GLCM and LBP, as well as the U-Net model, with the Jaccard and Dice indices used for performance comparison. These descriptors were employed to encode the textural information associated with the gland and stroma components. Specifically, the GLCM was used to calculate a co-occurrence matrix, providing information about the glands and stroma at various orientations, while the LBP was utilized to extract local intensity changes of the gland and stroma candidates.

The results demonstrated that the best accuracy was achieved using the GLCM at ($\delta = 1$, $\theta = 0$) and the LBP at $r = 1$, when employing an SVM classifier, yielding accuracy results of 99.0% and 95.1%, respectively, with

20-fold cross-validation. The combined hand-crafted model had an approximate training time of 2 minutes, both for training the model and performing image segmentation. However, despite the high classification accuracy, the hand-crafted models did not accurately delineate the segmentation boundary between the gland and stroma in the image patches. In contrast, the U-Net model, despite its longer training time of approximately 15 hours, achieved remarkable segmentation results even with a small training dataset. When tested on four histopathological images, the U-Net model outperformed the hand-crafted models, producing remarkable Jaccard and Dice index results, as presented in Table II. To improve readability, the fractional Jaccard Index and Dice similarity values were converted to percentage ratios.

Our results indicate significant differences between the hand-crafted (GLCM and LBP) and machine-driven (U-Net) approaches. Hand-crafted methods rely on predefined texture features, which can effectively capture specific patterns in prostate tissues. However, these methods may miss more complex patterns that U-Net, with its deep learning capabilities, can identify. U-Net's longer training time allows it to learn from data comprehensively, leading to superior segmentation quality but also requiring more computational resources. Therefore, the U-Net model's extended training time is crucial for achieving high accuracy, as longer training times enable the model to learn intricate patterns within the data, which improves segmentation results. However, this also raises concerns about computational costs and the potential for overfitting, particularly with smaller datasets. Extended work can focus on optimizing training processes and exploring techniques to reduce computational demands without sacrificing accuracy. Furthermore, investigating the generalizability of our approach to other types of cancer is a promising direction for future research. Preliminary experiments indicate potential applicability, which could enhance diagnostic tools across various cancer types.

Accordingly, results suggest that the U-Net model is the most effective for image segmentation and has the potential to assist pathologists in identifying early stages of Gleason grades of PCa, particularly in resource-constrained settings such as rural clinics, where limited clinical practitioners, infrastructure, and the unavailability of pathology services, along with missing historical data for many patients, are common challenges. This research could have significant clinical implications, including aiding pathologists in diagnosing prostate cancer more efficiently and accurately. Automated segmentation can reduce inter-observer variability and provide reliable diagnostic support, particularly in regions with limited healthcare resources. However, it is important to note that the limitation of the present work lies in the limited representational ability of hand-crafted features to handle significant variations in the anatomical shape's appearance. To complement domain knowledge and achieve better segmentation results, the utilization of other engineering hand-crafted models could be explored. Additionally, the proposed hand-crafted feature extraction methods were limited to grayscale images, and further investigation into other color models could potentially improve segmentation performance. Furthermore, exploring alternative training loss functions could be beneficial in training the U-Net model when dealing with limited data scenarios. Finally, the use of machine learning for prostate cancer identification offers significant benefits but may raises ethical concerns. Therefore, it is essential to use diverse and representative datasets, and involve healthcare providers in the development and training processes.

## V. CONCLUSIONS

This paper addresses the challenge of segmenting gland and stroma tissues in histopathological prostate images using a combination of handcrafted and semantic image segmentation techniques. The performance of various models, including hybrid features using GLCM and LBP, as well as the U-Net model, was evaluated and compared using the Jaccard and Dice indices. Remarkably, the U-Net model surpassed the handcrafted models in accurately segmenting gland and stroma regions. To enable automatic gland and stroma image segmentation, an online tool was developed based on the best trained model. Moving forward, enhancing performance by augmenting the image size in the training dataset and extending the model to handle whole-slide image segmentation are important future directions. Moreover, exploring architectural improvements, such as incorporating residual block connections and investigating alternative deep learning frameworks, can advance medical image segmentation beyond the standard U-Net architecture.

## CONFLICT OF INTEREST

The authors declare no conflict of interest.

## AUTHOR CONTRIBUTIONS

Feda Al beqain: Writing- Original draft preparation, Methodology, Software. Omar Al-Kadi: Supervisor, Conceptualization,Writing- Reviewing and Editing. All authors had approved the final version.